# Capacity Scaling of SDMA in Wireless Ad Hoc Networks

Marios Kountouris and Jeffrey G. Andrews

*Abstract*—We consider an ad hoc network in which each multi-antenna transmitter sends independent streams to multiple receivers in a Poisson field of interferers. We provide the outage probability and transmission capacity scaling laws, aiming at investigating the fundamental limits of Space Division Multiple Access (SDMA). We first show that super linear capacity scaling with the number of receive/transmit antennas can be achieved using dirty paper coding. Nevertheless, the potential benefits of multi-stream, multi-antenna communications fall off quickly if linear precoding is employed, leading to sublinear capacity growth in the case of single-antenna receivers. A key finding is that receive antenna array processing is of vital importance in SDMA ad hoc networks, as a means to cancel the increased residual interference and boost the signal power through diversity.

## I. INTRODUCTION

The last ten years have witnessed the transition of multi-antenna (MIMO) communication from a theoretical concept to a practical technique for enhancing performance of wireless networks. Point-to-point (single-user) MIMO communication promises large gains for both channel capacity and reliability, exhibiting linear capacity scaling with the minimum of the number of receive and transmit antennas [1], [2].

Fundamental information theoretic results advocate spatial sharing of the channel by the users. In space division multiple access (SDMA), the resulting multiuser interference is handled by the multiple antennas which in addition to providing per-link diversity also give the degrees of freedom necessary for spatial separation of the users. SDMA schemes, also known as multiuser MIMO, allow for a direct gain in multiple access capacity proportional to the number of transmit antennas. The capacity-achieving strategy for MIMO broadcast channels is dirty paper coding (DPC) [3], which is a theoretical pre-interference cancellation technique that requires perfect channel knowledge at the transmitter.

In this paper we are interested in the throughput gains that multiple antennas and multiuser MIMO may provide in uncoordinated ad hoc networks. Our focus is to determine the transmission capacity and outage probability scaling with the number of antennas as a function of network parameters. We aim at characterizing the fundamental limits of space division multiple access (SDMA) for both linear and non-linear precoding combined with various receive antenna array processing strategies. For that, we consider a network in which transmitters are randomly distributed on an infinite plane according to a 2-D homogeneous Poisson point process (PPP), and each transmitter attempts communication with multiple receivers, each located at a fixed distance away from it.

### A. Related work

There has been significant work on finding the transmission capacity of multi-antenna ad hoc networks in a Poisson field of interferers. In [4] the spectral efficiency of a single-hop ad hoc network with MIMO links and fixed density of interference is studied, and it is shown that the average per-link SINR increases with the number of receive antennas $N$ as $N^{2/\alpha}$ where $\alpha$ is the pathloss exponent. Based on tools from stochastic geometry, the authors in [5] provide capacity scaling laws when the $N$ antennas are used to cancel the $N-1$ strongest interferers. In [6] the performance of diversity-oriented receive processing (e.g. maximum ratio combining) is analyzed Both the above receive strategies achieve a sublinear density increase with $N$ (of order $N^{1-\frac{2}{\alpha}}$ and $N^{\frac{2}{\alpha}}$ respectively), which corresponds to only a logarithmic increase in per-link rate. Interestingly, it have been recently shown that using partial interference cancellation combined with array gain allows for linear density scaling with $N$ [7]. Spatial multiplexing MIMO techniques with perfect channel knowledge at the transmitter is analyzed in [8], showing the benefits from adaptive rate control.

### B. Contributions

In all prior work, each transmitter sends a single or multiple streams to only one receiver, whereas here we study multi-stream, multi-receiver transmission. A key finding is that the transmission capacity under dirty paper coding scales as $N^{1+\frac{2}{\alpha}}$. To the best of our knowledge, this is the first result in either centralized or ad hoc networks, where super linear capacity growth with the number of antennas is obtained. However, the interference boost due to aggressive linear multi-stream transmission techniques leads back to a sublinear scaling. We also highlight that receive antenna processing is able to restore the linear capacity scaling in ad hoc networks by utilizing a significant fraction of the available degrees of freedom to cancel the inter-node interference. A novel and general methodology for quantifying ad hoc capacity is developed when the total interference consists of inter-user (SDMA) and Poisson field interference. Interestingly, this framework can be used to analyze the performance of SDMA with limited feedback.

## II. NETWORK MODEL AND PRELIMINARIES

We consider a network in which the transmitting nodes are distributed according to a stationary Poisson point process

M. Kountouris and J. G. Andrews are with the University of Texas at Austin. This research has been supported by the DARPA IT-MANET program.

(PPP) with intensity $\lambda$ on the plane. This is a realistic model assuming that the transmitting nodes in the network are randomly and independently located and do not cooperate. The signal strength is subject to pathloss attenuation model, $|\cdot|^{-\alpha}$ for a distance $d$ with exponent $\alpha > 2$ as well as small scale unit-mean Rayleigh fading. The interfering nodes constitute a marked PPP, denoted as $\Pi = \{X_i, I_i\}$, where each $X_i \in \mathbb{R}^2$ is the location of the $i$-th interfering transmitter, and with marks $I_i$ that denote the fading factors on the power transmitted from the $i$-th node and then received by a typical receiver.

Let all nodes transmit with the same power $\rho$ and $H_0$ be the signal fading between a typical receiver and its intended transmitter, labeled $T_0$. Assuming that all receivers are located at a fixed distance $D$ away from their transmitter, the resulting signal-to-interference-plus-noise ratio (SINR) is given by

$$\text{SINR} = \frac{\rho H_0 D^{-\alpha}}{\rho \sum_{i \in \Pi(\lambda)} I_i |X_i|^{-\alpha} + \eta} \quad (1)$$

where $\eta$ is the background noise power and $I_i$ is the fading coefficient from the $i$-th interferer (with $H_0, I_1, I_2, \ldots$ independent). In contrast to the above related work, we explicitly consider noise, whose effect is important in the power-limited regime.

*A. MIMO SDMA Channel Model*

We consider a network in which each transmitter with $M$ antennas communicates simultaneously with $|\mathcal{K}| = K \leq M$ receivers, each with $N$ receive antennas. In this point-to-multipoint context, each of the $K$ streams sent by the transmitter contains a separate message destined to different receivers. By the stationarity of the Poisson process, it is sufficient to analyze the the performance of a typical TX - multi-RX link, which we refer to as $\text{TX}_0$ and $\text{RX}_0^{(k)}$, for $k = 1, \ldots, K$. From the perspective of each typical receiver, the set of interferers (which is the entire transmit process with the exception of $\text{TX}_0$) also form a homogeneous PPP due to Slivnyak's Theorem [9].

The received signal $\mathbf{y}_k$ at reveiver $k \in \mathcal{K}$ is given by

$$\mathbf{y}_k = \sqrt{\frac{P}{M}} D^{-\alpha/2} \mathbf{H}_{0k} \mathbf{x}_k + \sqrt{\frac{P}{M}} \sum_{i \in \Pi} |X_i|^{-\alpha/2} \mathbf{H}_{ik} \mathbf{x}_i + \mathbf{n} \quad (2)$$

where $\mathbf{H}_{0k} \in \mathbb{C}^{N \times M}$ is the channel between $T_0$ and receiver $k$, $\mathbf{H}_{ik} \in \mathbb{C}^{N \times M}$ is the channel between receiver $k$ and interfering transmitters $T_i$, $\mathbf{x}_k$ is the normalized transmit signal vector, and $\mathbf{n}$ is complex additive Gaussian noise. Unless otherwise stated, we assume that each transmitter has perfect knowledge of the channels (CSI) of its intended receivers, and each receiver has perfect knowledge of its own channel matrix. For exposition convenience, we drop the index 0 as all the subsequent analysis is performed on the above typical link.

*B. Key Performance Metrics*

A primary performance metric of interest in an uncoordinated, random access ad hoc network is the outage probability $\mathcal{P}_{\text{out}}$ with respect to a pre-defined target SINR $\beta$. A message is successfully decoded if $\text{SINR} \geq \beta$ and for a failure probability $\epsilon$, we have

$$\mathcal{P}_{\text{out}} = \mathbb{P}(\text{SINR} \leq \beta) \leq \epsilon. \quad (3)$$

The SINR statistics are a function of the interferers density $\lambda$ and $\mathcal{P}_{\text{out}}$ is clearly an increasing function of $\lambda$. In this paper, multiple streams are sent by each transmitter, thus different SINR statistics may be seen on the different streams, resulting on a per-stream outage constraint $\epsilon_k$. If the target SINR on stream $k$ is given by $\beta_k$, the maximum density of concurrent (single-stream) transmissions $\lambda_\epsilon$ per m$^2$ such that (3) is satisfied, is defined as

$$\lambda_\epsilon = \max\{\lambda : \mathbb{P}(\text{SINR}_k \leq \beta_k) \leq \epsilon_k, \forall k\}. \quad (4)$$

In our analysis, we assume that the SINR statistics are identical for each receiver. In general, if target SINR and outage constraints are specified for each stream (subchannel), then the weakest stream will become the limiting factor in the optimal contention density, and hence the optimal area spectral efficiency. Assuming transmission at the Shannon target rate $b = \log_2(1 + \beta)$ bps/Hz, the area spectral efficiency is defined as

$$\mathcal{C}_\epsilon = K \lambda_\epsilon (1 - \epsilon) b \quad \text{bps/Hz/m}^2 \quad (5)$$

and depends on the number $1 \leq K \leq M$ spatial streams sent by each source node.

### III. DIRTY PAPER CODING

In this section, we derive upper and lower capacity bounds when dirty paper coding is employed. Despite being capacity-achieving for MIMO broadcast channels, DPC is not necessarily optimal in ad hoc networks, which can be considered as compound interference channels. However, joint optimization of all precoding matrices is a challenging task, which requires global CSI of the instantaneous channel conditions of all transmitting nodes (interferers). Furthermore, such optimization induces heterogeneous and not necessarily Poisson point processes for which closed-form results are hard or impossible to obtain.

We first derive two general bounds on the outage probability when the channel gain follows a chi-square ($\chi^2$) distribution. Let $\mathcal{L}_Y$ denote the Laplace transform of the probability density function (pdf) of the interference $Y$, defined as $\mathcal{L}_Y(s) = \int_0^\infty e^{-sy} f_Y(y) dy = \mathbb{E}\left[e^{-sY}\right]$. The Laplace transform of the noise is defined as $\mathcal{L}_N(s) = e^{\eta s}$

**Lemma 1:** *The success probability in a random access network, where the channel fading follows a chi-square distribution with $2d$ degrees of freedom, i.e. $H_0 \sim \chi^2_{(2d)}$, is upper bounded by*

$$\mathbb{P}(\text{SINR} \geq \beta) \leq \mathcal{L}_Y(s) \mathcal{L}_N(s/\rho) \quad (6)$$

*with $s = \frac{\beta D^\alpha}{4d}$.*

*Proof:* See [10]. ∎

The above lemma provides a large-deviation bound on the success probability by bounding the tail of the received signal (i.e. a $\chi^2$ random variate). In other words, a concentration inequality of a central $\chi^2$ variate is exploited to upper bound the probability the received signal power falls away from its mean $d$. Although Lemma 1 captures the essential capacity scaling, the following result ought to be used for more accurate bounds on the success probability.

***Lemma** 2:* The outage probability in a random access network where $H_0 \sim \chi^2_{(2d)}$ is bounded by

$$\mathcal{A}_d(ck\beta D^\alpha) \leq \mathbb{P}\left(\text{SINR} \leq \beta\right) \leq \mathcal{A}_d(k\beta D^\alpha) \quad (7)$$

where $\mathcal{A}_d(\zeta) = \sum_{k=0}^d \binom{d}{k}(-1)^k \mathcal{L}_Y(\zeta)\mathcal{L}_N(\zeta/\rho)$ and $c = (d!)^{-1/d}$.

*Proof:* See [10]. ∎

The above lemma relies on tightest known bound for the distribution function of $\chi^2$ distribution, and is more accurate yet more involved than (6).

### A. DPC with Multi-antenna Receivers

The exact received signal power at each receiver is hard to derive since no closed-form solution for the optimal DPC transmit covariance matrices is known. However, an upper bound on the SINR can be derived by bounding the received signal power and considering $H_0 \sim \mu^2_{max}(\mathbf{H}_0^H \mathbf{H}_0)$, where $\mu_{max}$ denotes the maximum eigenvalue of $\mathbf{H}_0^H \mathbf{H}_0$. Since the distribution (cdf) of the square of the maximum singular value of the user channel is involved (sums of exponentials and Laguerre polynomials), closed-form expressions yield little intuition. For that, the largest squared singular value is upper bounded by $\mu^2_{max} \leq \|\mathbf{H}_0\|^2_F \sim \chi^2_{(2MN)}$. The marks of the PPP interference are sums of the interference from the $M$ independent messages transmitted by each interfering node, i.e. the sum of $M$ independent $\chi^2_{(2)}$ (due to the independence of the channels and the precoders of interfering transmitters with the receive filters). Thus, $I_k \sim \chi^2_{(2M)}$ and the Laplace transform is given by $\mathcal{L}_Y(s) = e^{-\lambda \beta^{2/\alpha} D^2 \mathcal{I}_M}$ [8] where

$$\mathcal{I}_M = \frac{2\pi}{\alpha} \sum_{m=0}^{M-1} \binom{M}{m} B(m+2/\alpha, M-(m+2/\alpha)) \quad (8)$$

with $B(a,b) = \int_0^1 t^{a-1}(1-t)^{b-1} dt$ being the Beta function. Based on (8), Lemma 1 applies as follows:

***Proposition** 1:* The maximum contention density of an ad hoc network in which each transmitter communicates with $M$ receivers using dirty paper coding, is upper bounded by

$$\lambda_{\text{DPC}} \leq \frac{(4MN)^{2/\alpha}}{\mathcal{I}_M \beta^{2/\alpha} D^2}\left[-\log(1-\epsilon) + \frac{\eta\beta D^\alpha}{4MN\rho}\right]. \quad (9)$$

The second term in (9) captures the effect of background noise and has a additive contribution on the contention density. Note that the noise effect falls off to zero for large number of transmit/receive antennas.

Since $\mathcal{I}_M \sim \pi \Gamma(1-2/\alpha) M^{2/\alpha}$ for large $M$, it can be easily shown that:

***Lemma** 3:* The transmission capacity employing dirty paper coding scales super linearly with the number of antennas, i.e.

$$\mathcal{C}_{\text{DPC}} = O(MN^{2/a}) \quad (10)$$

For $M = N$, the ASE scales as $\mathcal{C}_{\text{DPC}} = O(N^{1+2/a})$. As shown in [7], a linear scaling of $O(N)$ can be achieved in ad hoc SIMO channels using partial zero-forcing (PZF), thus orderwise we have $\frac{\mathcal{C}_{\text{DPC}}}{\mathcal{C}_{\text{PZF}}} = O(N^{2/\alpha})$. In order words, the interference pre-substraction capability of DPC allows for $N^{2/\alpha}$ more concurrent transmissions (streams) per unit area in a random access network as compared to single-stream MIMO communications.

For small outage constraints $\epsilon$, by expanding $\mathcal{L}_Y(s)$ using first order Taylor series around zero we have that

***Lemma** 4:* The optimal contention density $\lambda^{\text{mimo}}_{\text{DPC}}$ when DPC precoding is employed to $M$ multi-antenna receivers is given by

$$\lambda^{\text{mimo}}_{\text{DPC}} = \frac{\mathcal{F}_{MN}\epsilon}{\mathcal{I}_M \beta^{2/\alpha} D^2} e^{-\frac{\eta\beta D^\alpha}{\rho}} \quad (11)$$

where

$$\mathcal{F}_{MN} = \left[\sum_{k=0}^{MN-1}\sum_{j=0}^k \binom{k}{j}\left(\frac{\eta}{\rho}\right)^{k-j}\frac{1}{j!}\prod_{m=0}^{j-1}(m-2/\alpha)\right]^{-1}. \quad (12)$$

*Proof:* See [10]. ∎

Note that for large number of antennas, $\mathcal{F}_{MN}$ increases as $O((MN)^{\frac{2}{\alpha}})$, which is consistent with the scaling given in Lemma 3. Similarly, applying Lemma 2, the following bounds can be derived:

***Proposition** 2:* If DPC is employed, the maximum density under an outage constraint $\epsilon$ is lower bounded by

$$\frac{\left(\epsilon - (1-e^{\frac{\eta\zeta}{\rho}})^d\right)\mathcal{S}^{-1}_{d,1}}{\mathcal{I}_M \beta^{2/\alpha} D^2} \leq \lambda_{\text{DPC}} \leq \frac{\left(\epsilon - (1-e^{\frac{\vartheta\eta\zeta}{\rho}})^d\right)\mathcal{S}^{-1}_{d,\vartheta}}{\vartheta^{2/\alpha}\mathcal{I}_M \beta^{2/\alpha} D^2} \quad (13)$$

with $d = MN$, $\zeta = \beta D^\alpha$, and $\vartheta = \Gamma(d+1)^{-1/d}$.

$$\mathcal{S}_{d,\vartheta} = \sum_{n=1}^d \binom{d}{n}(-1)^{n+1}n^{2/\alpha}e^{\frac{\vartheta\eta\zeta}{\rho}} \quad (14)$$

For asymptotically large $d$, $\Gamma(d+1)^{\frac{2}{\alpha d}} \sim d^{2/\alpha}$ (using Stirling's formula) and $\mathcal{I}_M \sim M^{2/\alpha}$. The asymptotic capacity scaling $\mathcal{C}_{\text{DPC}}$ depends on the scaling of $\mathcal{S}_{MN}$ for large number of antennas.

### B. DPC with Single-antenna Receivers

When each transmitter communicates with $M$ single-antenna receivers, no receive antenna processing/combining can be performed; therefore $H_0$ is at best distributed as $\chi^2_{(2M)}$, whereas the interference marks remain unchanged, i.e. $I_i \sim \chi^2_{(2M)}$ Following the analysis in [6], we can show that

***Proposition** 3:* Transmission to $M$ single-antenna receivers (MISO) using DPC precoding results in a maximum contention density $\lambda^{\text{miso}}_{\text{DPC}}$ of

$$\lambda^{\text{miso}}_{\text{DPC}} = \frac{\mathcal{F}_M \epsilon}{\mathcal{I}_M \beta^{2/\alpha} D^2} e^{-\frac{\eta\beta D^\alpha}{\rho}}. \quad (15)$$

where $\mathcal{F}_M$ is given in (12).

Since for large $M$, $\frac{\mathcal{F}_M}{\mathcal{I}_M} = O(1)$, the transmission capacity exhibits linear scaling, i.e. $\mathcal{C}^{miso}_{\text{DPC}} = O(M)$. The lack of receive antennas leads to no receive diversity or interference cancellation gain, thus the per-user outage probability is of order $O(1)$. The linear scaling in the ASE with the number of transmit antennas is mainly due to the fact that $M$ concurrent streams per transmitter are sent.

## IV. LINEAR PRECODING WITH ANTENNA COMBINING

Since the complexity of dirty paper coding is very high, linear precoding has attracted wide attention as a low complexity (but suboptimal) technique with complexity roughly equivalent to point-to-point MIMO systems. Since linear precoding is able to transmit the same number of data streams as a DPC-based system, it therefore achieves the same multiplexing gain in a MIMO broadcast channel, but incurs a power offset relative to DPC. In this section, we aim at deriving the achievable throughput of SDMA transmission when linear precoding techniques are employed.

### A. Zero-forcing Beamforming

When zero-forcing beamforming (ZFBF) is employed, the beamforming vectors are chosen such that no inter-user (SDMA) interference is experienced at any of the receivers.

*1) ZFBF with Multi-antenna Receivers:* For $M \geq KN$ with $N > 1$, the precoding matrix $\{\mathbf{W}_j\}_{j=1}^{K}$ is chosen such that at each receive antenna $n$, the zero inter-stream interference constraint imposes: $\mathbf{h}_{k,n}\mathbf{w}_{j,l} = 0, \forall j \neq k, \forall n, l \in [1, N]$ and $\mathbf{h}_{k,n}\mathbf{w}_{k,l} = 0, \forall l \neq n$. Therefore, the the effective channel gain at each receive antenna is given by $|\mathbf{h}_{k,n}\mathbf{w}_{k,n}|^2$, which follows a $\chi^2_{2(M-KN+1)}$ distribution. Similarly to Proposition 3, we can show that the maximum contention density in the small outage constraint regime is given by

$$\lambda_{\text{ZF}} = \frac{\mathcal{F}_c \epsilon}{\mathcal{I}_{KN} \beta^{2/\alpha} D^2} e^{-\frac{\eta \beta D^\alpha}{\rho}}. \quad (16)$$

with $c = M - KN + 1$. Therefore, the capacity scales as $\mathcal{C}_{\text{ZF}} = O(\frac{(M-KN+1)^{2/\alpha}}{(KN)^{2/\alpha-1}})$, which equals $O((KN)^{2/\alpha})$ for $M = KN$.

If $N > M$, the extra degrees of freedom available at the receiver side can be exploited to eliminate the inter-node interference (e.g. employing a zero-forcing linear receive filter). In that case, the received signal power is distributed as $H_0 \sim \chi^2_{2(N-M+1)}$, whereas the interference marks are sums of $M$ $\chi^2_{(2)}$ random variables. This results in $\mathcal{C}_{\text{ZF}} = O(M(\frac{N-M+1}{M})^{\frac{2}{\alpha}}) = O(M^{1-2/\alpha}) < O(N^{1-2/\alpha})$. The fact that multi-stream transmission boosts the interference coming from interfering transmitters, zero-forcing linear processing fails to provide capacity scaling $N^{1-2/\alpha}$ with the number of receive antennas, as in [5].

*2) ZFBF with Receive Antenna Selection:* Consider now a system that performs antenna selection at the receiver side, i.e. each destination node selects its best receive antenna. The channel vector $\mathbf{h}_k$ fed back from receiver $k$ to the transmitter corresponds to the antenna that has the best instantaneous channel (best among $N$ i.i.d. channel vectors). Therefore, $H_0 \sim \max_{1 \leq n \leq N} \mathbf{h}_k^{(n)}$, while the interference marks remain the same as in Section III. For a such network, as $F_{H_0}(x) = (1 - e^{-x})^N = \sum_{n=1}^{N} \binom{N}{n}(-1)^{n+1} e^{-nx}$, the maximum contention density (for $\epsilon \to 0$) is given by

$$\lambda_{\text{ZF}}^{as} = \frac{\epsilon}{\mathcal{S}'_N \mathcal{I}_M \beta^{2/\alpha} D^2} e^{-\frac{\eta \beta D^\alpha}{\rho}} \quad (17)$$

where $\mathcal{S}'_N = \sum_{n=1}^{d} \sum_{j=1}^{n} \binom{n}{j}\binom{d}{j}\left(\frac{\eta}{\rho}\right)^{(n-j)}(-1)^{j+1} j^{2/\alpha}$ and $\mathcal{I}_M$ is given by (8).

Therefore, the capacity scaling depends on the scaling of $\mathcal{S}_N$, i.e. $\mathcal{C}_{\text{ZF}}^{as} = O(\mathcal{S}_N^{-1} M^{1-\frac{2}{\alpha}})$. For $M = N$, we have that $\mathcal{C}_{\text{ZF}}^{as} = \Theta(M)$, due to the fact that selection improves the typical channel without amplifying interference. Since order statistics (due to selection) provides an $M^{2/\alpha}$-fold increase of the received signal power, a linear capacity growth with the number of antennas can be achieved.

*3) ZFBF with Single-antenna Receivers:* The beamforming vector of receiver $k$, denoted as $\mathbf{w}_k$, is chosen to be orthogonal to the channel vectors of all other intended receivers, i.e. $\mathbf{h}_j \mathbf{w}_k = 0, \forall j \in \mathcal{K}, j \neq k$. By construction the distribution of $H_0 = |\mathbf{h}_k \mathbf{w}_k|^2$ is $\chi^2_{(2)}$, whereas the interference marks $I_i$ remain gamma distributed, i.e. $I_i \sim \chi^2_{(2M)}$. The following proposition characterizes the performance of ZFBF in MISO SDMA ad hoc networks:

***Proposition 4:*** *For a random access wireless network in which the transmitters spatially multiplex $M$ single-antenna receivers using ZFBF, the maximum density under an outage constraint $\epsilon$ is given by*

$$\lambda_{\text{ZF}} = \frac{-\log(1-\epsilon)}{\mathcal{I}_M \beta^{2/\alpha} D^2} + \frac{\eta \beta^{1-\frac{2}{\alpha}} D^{\alpha-2}}{\rho \mathcal{I}_M} \quad (18)$$

*where $\mathcal{I}_M$ is given in (8).*

*Proof:* See [10] ■

Thus, for large $M$, the interference power increases with $M^{2/\alpha}$, whereas the signal power does not increase with the number of antennas, resulting in asymptotic ASE scaling of $\mathcal{C}_{\text{ZF}} = O(M^{1-2/\alpha})$. Note that the same orderwise capacity scaling is achieved by interference-aware beamforming [5], where a $N$-dimensional received array is used to cancel the nearest $N-1$ interferers, under the assumption that the receiver has knowledge of the interferers' channels. Furthermore, it can be shown that regularized channel inversion (MMSE precoding) provides the same $O(M^{1-2/\alpha})$ scaling, however higher SINR target $\beta$ per user stream may be achieved for the same outage constraint $\epsilon$.

### B. Block Diagonalization

If $M \geq KN$, SDMA inter-user interference can also be eliminated by using block diagonalization (BD). The precoding matrix is chosen to be $\mathbf{H}_k \mathbf{W}_j = \mathbf{0}$, for $k \neq j$, thus converting the system into $K$ parallel MIMO channels with effective channel matrices $\mathbf{G}_k = \mathbf{H}_k \mathbf{W}_k$. Since the network is interference-limited, equal power allocation is asymptotically (in SNR) optimal, thus an upper bound on the SINR under BD can be found by considering that the received power is equal to the square of the maximum singular value $\mu_{max}^2(\mathbf{G}_k \mathbf{G}_k^H)$ scaled by the path loss and the transmit power. Since $\mathbf{G}_k$ is a Wishart matrix with $N \times (M - (K-1)N)$ degrees of freedom and $\mu_{max}^2 \leq \|\mathbf{G}_k\| \sim \chi^2_{(2(NM-(K-1)N^2))}$ [11], applying Theorem 1 in [6], it can be easily shown that

*Proposition* 5: For small outage constraints, the maximum density using BD is upper bounded by

$$\lambda_{BD} \leq \frac{\mathcal{F}_r \epsilon}{\mathcal{I}_K \beta^{2/\alpha} D^2} e^{-\frac{\eta \beta D^\alpha}{\rho}}. \quad (19)$$

where $r = NM - (K-1)N^2$, $\mathcal{F}_r$ is given by (12), and $\mathcal{I}_K$ is given by (8)

*Proof:* See [10]. ∎

For large number of transmit/receive antennas, we have that $\mathcal{F}_r \sim (NM - (K-1)N^2)^{2/\alpha}$, thus assuming $M = KN$, we have that $\lambda_{BD} \leq O(N^{\frac{4}{\alpha}} K^{-\frac{2}{\alpha}})$. Therefore, the capacity scaling when block diagonalization is employed scales as $\mathcal{C}_{BD} \leq O(N^{4/\alpha} K^{1-2/\alpha})$, which is a decreasing function with $K$. This implies that capacity growth is maximized if $K = 1$ receiver is served, leading to a super linear scaling for $\alpha < 4$. The scaling law is orderwise equivalent to $N \times N$ eigenbeamforming transmission (spatial multiplexing) with interference cancellation of the $N - 1$ closest interferers.

## V. NUMERICAL RESULTS AND DISCUSSION

In this section, we assess the performance of SDMA ad hoc networks with default parameters $D = 10$m, $\epsilon = 0.1$, $\alpha = 4$, and $\beta = 3$ as a means to verify our theoretical analysis. We adopt the practically relevant assumption $M = N$.

In Figure 1 the DPC transmission capacity is compared with the derived upper bounds (cf. (9) and (13)) and the analytical performance versus the number of transmit antennas in the interference-limited regime. As predicted, dirty paper coding exhibits a super linear scaling behavior with the number of antennas, also captured by the derived bounds. The tightness of the upper bounds depends on the pathloss exponent $\alpha$ and $M$, being tighter for $\alpha$ decreasing. Furthermore, for small outage constraint, (11) accurately predicts the SDMA transmission capacity of dirty paper coding. Note also that substantial gains appear even when only a few streams are transmitted.

In Figure 2 we verify the linear scaling of DPC with single-antenna receivers as well as the sublinear capacity behavior of linear precoding. It should be noticed that diversity-oriented receive processing combined with linear transmit processing is not sufficient to achieve linear scaling, as compared to DPC even with one receive antenna. This is mainly to the fact multi-stream transmissions boost the aggregate interference seen at a typical receiver, resulting in similar spectral efficiency performance as in the point-to-point MIMO case.

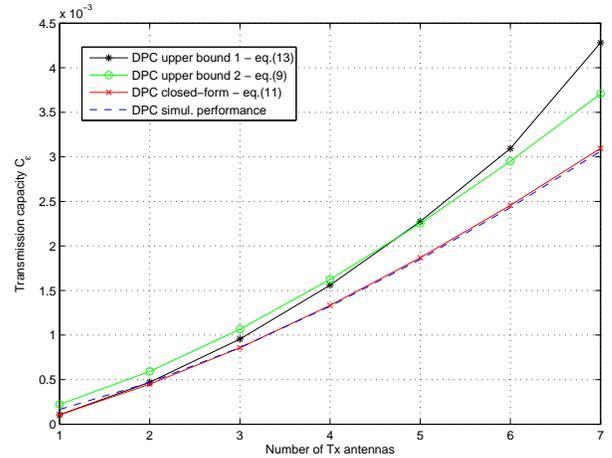

Fig. 1. Capacity scaling $\mathcal{C}_{\text{DPC}}$ of dirty paper coding versus the number of transmit antennas ($M = N$) for $\alpha = 4$.

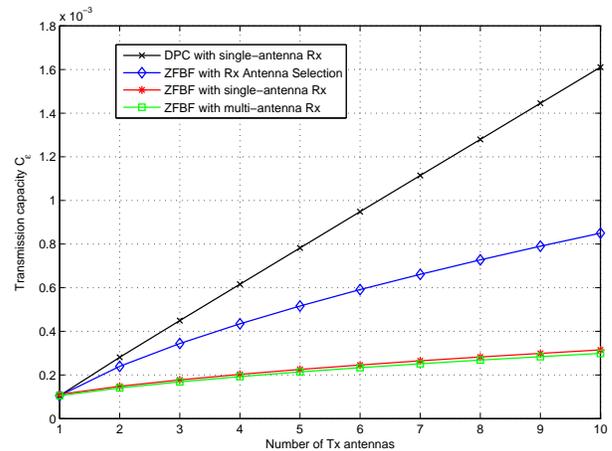

Fig. 2. Transmission capacity versus the number of antennas for different SDMA precoding techniques and $\alpha = 4$.